\documentclass[12pt]{article}
\usepackage{amsmath,amssymb,physics,booktabs}
\usepackage{graphicx}
\usepackage{bm}
\usepackage{geometry}
\usepackage{hyperref}
\usepackage{authblk}
\usepackage{enumitem}
\usepackage{physics}
\usepackage{tikz}
\usepackage{hyperref}
\usepackage{xcolor}
\title{Quantum-like Architecture of the Social Atom: Mental Marker Coherence and the Triple-Resonance Mechanism of Cognitive Activation}
\author{Andrei Khrennikov$^{1,*}$, Christian Aspalter$^{2}$, Likan Zhan$^{3}$} 
\date{}

\begin{document}
\maketitle
\noindent
$^{1,*}$Center for Mathematical Modeling in Physics and Cognitive Sciences\\
Linnaeus University, V\"axj\"o, SE-351 95, Sweden\\
$^{2}$ Department of Social Sciences, Beijing Normal-Hong Kong Baptist University (BNBU), Zhuhai, China\\
$^{3}$ School of Psychology and Cognitive Science, Beijing Language and Culture University (BLCU), Beijing, China\\
*Corresponding author email: Andrei.Khrennikov@lnu.se

\begin{abstract}
Social Laser Theory (SLT) provides a rigorous quantum-like mathematical formalism for modeling collective social activation. While its macroscopic effects mimic laser physics, the fundamental driver of the "lasing" regime resides in the internal state dynamics of the "social atom." In this paper, we propose a quantum-like architecture for the individual cognitive system, formalizing the social atom as a composite quantum system where mental markers—spanning cognitive, affective, and valence dimensions—are represented as internal degrees of freedom in a complex Hilbert space. We show that social activation is not a classical linear response but is governed by a Triple-Resonance Mechanism: the simultaneous quantum-like matching of semantic topic, energetic impact, and valence stance between the agent's internal state and the informational field. We detail how the repeated processing of quantized informational units ("infons") leads to a state of cognitive population inversion. We further establish a link to Human Entanglement Theory (HET), suggesting that the emergence of coherent informational fields is rooted in the phase alignment of these internal mental markers across a population. By situating SLT within the broader framework of Social Quantum Field Theory (SQFT), this paper provides a neuro-cognitive foundation for collective phenomena such as emotional contagion and rapid narrative amplification. Our results suggest that the "Social Laser" is a macroscopic manifestation of quantum-like effects across the human cognitive and neuro-social architecture.
\end{abstract}

Social Laser Theory (SLT) provides a rigorous quantum-like mathematical formalism for modeling collective social activation. While its macroscopic effects mimic laser physics, the fundamental driver of the lasing regime resides in the internal state dynamics of the social atom. In this paper, we propose a quantum-like architecture for the individual cognitive system, formalizing the social atom as a composite quantum system where mental markers—spanning cognitive, affective, and valence dimensions - are represented as internal degrees of freedom in a complex Hilbert space. We show that social activation is not a classical linear response but is governed by a Triple-Resonance Mechanism: the simultaneous quantum-like matching of semantic topic, energetic impact, and valence stance between the agent's internal state and the informational field. We detail how the repeated processing of quantized informational units (infons) leads to a state of cognitive population inversion. We further establish a link to Human Entanglement Theory (HET), suggesting that the emergence of coherent informational fields is rooted in the phase alignment of these internal mental markers across a population. By situating SLT within the broader framework of Social Quantum Field Theory (SQFT), this paper provides a neuro-cognitive foundation for collective phenomena such as emotional contagion and rapid narrative amplification. Our results suggest that the Social Laser is a macroscopic manifestation of quantum-like effects across the human cognitive and neuro-social architecture.

{\bf keywords:} quantum-like modeling; social laser theory; mental markers; triple-resonance mechanism; Hilbert space; social quantum field theory; cognitive activation; phase coherence; human entanglement; population inversion

\section{Introduction: From Neuro-Cognitive Mental Markers to Socio-Political Force Fields}

The rapid development of quantum information theory---often described as the \textit{second quantum revolution}---has stimulated a profound shift in our understanding of complex systems. This evolution has sparked renewed interest in the foundational role of quantum theory and its potential applicability across neuroscience and cognitive science. A significant outcome of this progress is the emergence of \textit{quantum-like modeling} (QLM), an interdisciplinary research program that applies the probabilistic and mathematical structure of quantum theory to complex systems in neurobiology, cognition, decision theory, economics, and the social sciences \cite{INT, framingP, Aerts3, AGS, ATM1, ATM3, Bruza, Busemeyer5, Asano1, QL3, Bagarell2, Bagarello4, P&B, PL4, Zeit, OK20, OK21}. This article specifically explores how these quantum-like effects manifest across the human cognitive architecture, from neural signal processing to the higher-order organization of collective mental states.

\subsection{The Social Atom as a Neuro-Cognitive Qubit}
Within this broader program, \textit{Social Laser Theory} (SLT) \cite{laser1, laser2, laser5, laser6, laser12, laser13} has been proposed as a robust framework for modeling large-scale social activation phenomena by drawing inspiration from the physics of stimulated emission. However, the efficacy of the ``social laser'' depends fundamentally on the internal architecture of the \textit{social atom}---the individual information-processing agent (person). In our framework, we move beyond viewing the person as a classical rational actor; instead, we formalize the individual as a composite quantum system. Here, \textit{mental markers}---spanning cognitive, affective, and valence-based dimensions---represent internal degrees of freedom within a complex Hilbert space.

These mental markers function as neuro-cognitive energy levels, reflecting a motivational readiness for participation in collective action. The transition between these levels is triggered by discrete informational excitations called \textit{infons}. Unlike classical stimuli, infons interact with the social atom through a \textit{Triple-Resonance Mechanism}, matching the agent’s internal mental markers in a process that mirrors the absorption and emission of photons in physical systems. Infons with matching markers lead to an excitation of energy levels related to pro-collective action potentials.

\subsection{Parallelity, Orthogonality, and Neural Alignment}
The degree of alignment between an agent’s internal state and the incoming information is captured by the semantic overlap coefficient $\Gamma_{kj}$. In quantum-neuroscience terms, this represents the absence of \textit{orthogonality} (opposition) between the agent’s state and the informational field. When this ``social parallelity'' is high, the transition probability $P_{0 \to 1}$ increases, reflecting a neuro-cognitive alignment that facilitates message acceptance. Conversely, as the mental markers of the agent and the infon approach orthogonality, the probability of rejection, counter-mobilization, and conflict dominates.

Crucially, these cognitive states do not respond to stimuli in a linear causal way. Instead, state-transitions exhibit critical \textit{quantum-like vanishing points} and \textit{acceleration points}, mirroring the developmental dynamics of quantum vortices in physical quantum systems \cite{Aspalter5, Aspalter7}. This suggests that the brain’s ``ready-state'' for social mobilization is a non-linear accumulation of mental-marker excitations, eventually leading to a state of \textit{population inversion} within the social group, where the majority of agents are primed for coherent activation.

\subsection{Foucauldian Fields and Human Entanglement}
To understand the ``tuning'' of these mental markers, we must look beyond direct signaling. Causality in human action is a complex bridge between direct informational impact and broader \textit{Human Entanglement Theory} (HET), \textit{Mental Entanglement Theory} (MET), and \textit{Social Quantum Field Theory} (SQFT) \cite{Aspalter2, Aspalter3, Aspalter5, Aspalter7, Asano1, Asano2, E1, E2, E3, Bruza, QL3, Behti}. These entanglements represent probabilistic causal relationships inhabiting the bosonic Fock space ($\mathcal{H}_{field}$) of the social system.\footnote{We also point to the works Eberly et al. \cite{Eberly}, Bardella et al. \cite{Bardella1,Bardella2}, Buice \cite{Buice1,Buice2}, Gentili \cite{Gentili}  focused on developing field models—both classical and quantum—of neural activity in the brain.}

We propose that these entanglements function as \textit{socio-political mental force fields}. To define their nature, we draw on \textit{Foucauldian power relations} \cite{Foucault, Gramsci, Aspalter2, Aspalter3, Aspalter5, Aspalter7}. In our neuro-cognitive model, power is the ``Social Field'' that pre-configures the social atom’s internal state. Much like a particle in a field, the agent’s mental markers are tuned by a web of norms, laws, and social expectations---a \textit{Foucauldian Field}---making certain transitions statistically more likely even before a specific infon arrives. 

\subsection{Signaling and Self-Emergence in the Bosonic Space}
This field-based perspective implies that social causality is a manifestation of varied entanglements, ranging from direct signaling to ``self-emergent'' forces that travel through both conscious and sub-conscious neural pathways. We distinguish between signaling-based entanglements and non-signaling, self-emergent forces where agents follow fashion or politics through an internal extrapolation of societal and governmental rules \cite{Freud, Weber, Aspalter1}.

Furthermore, we identify the role of \textit{atmospheric interpellation}---the causally amplifying influence of dominant public sentiment \cite{Aspalter5}. These collective sentiments act as macroscopic mental force fields that inhabit the bosonic Fock space. As Friedrich Hayek \cite{Hayek} emphasized regarding the ideological climate, and Friedrich Wieser \cite{Wieser} depicted as the internal powers of the state, these fields are individually anchored yet collectively coherent. Even Pericles noted these intangible forces that penetrate the human psyche, dominating areas of action untouched by direct policy. In the context of this Special Issue, we argue that these forces represent non-signaling-based, self-emergent human entanglements that organize the quantum-like probability distributions of human cognition and collective decision-making.

\subsection{Structural Analogy Between Physical and Social Lasers}

\begin{table}[h]
\centering
\begin{tabular}{lll}
\toprule
\textbf{Physical Laser} & \textbf{Social Laser Theory} & \textbf{Interpretation} \\
\midrule
Atoms & Social atoms (agents) & Information-processing individuals \\
Photon & Infon & Quantum of social information \\
Energy levels & Social energy states & Readiness for social action \\
Population inversion & Social excitation & Large activated population \\
Coherent EM field & Coherent infon field & Structured discourse environment \\
Phase coherence & Discursive alignment & Messages reinforce each other \\
Photon statistics & Infon statistics & Regimes of communication dynamics \\
\bottomrule
\end{tabular}
\caption{Basic analogy between physical lasers and Social Laser Theory}
\end{table}

In quantum optics, the radiation field produced by a laser operating above
threshold is well approximated by a \textit{coherent state} of the
electromagnetic field. Such states exhibit classical-like behavior with a
well-defined amplitude and phase and are characterized by Poissonian photon statistics. The crucial feature enabling laser amplification is
\textit{phase inheritance}: photons generated through stimulated emission
adopt the phase of the existing field, thereby reinforcing the collective
radiation mode.

An analogous mechanism can be proposed for the dynamics of social
communication. In SLT, the informational environment is modeled as a quantum information field 
 composed of discrete informational quanta -- \textit{infons}.
When communication signals become aligned in meaning, emotional tone,
and symbolic framing, the resulting informational field may approach a
state of partial coherence in which successive messages reinforce one
another rather than interfere destructively. Emotional and ideological coherence is the outcome of greater degrees of inter-agential alignment (parallelity), and greater number of repetition and/or higher intensity (amplitudes) of emotional groundedness.

A key element of this mechanism is the structure of informational quanta
themselves. In Section~2.3 we introduce the notion of
\textbf{affective markers}, minimal informational units capable of
triggering emotional resonance across individuals. Unlike complex
propositional statements, affective markers often carry little semantic
content but possess strong emotional valence. Empirical research shows
that emotionally charged signals propagate efficiently through social
networks and digital media environments, producing cascades of collective
attention and behavioral contagion
\cite{Hatfield,centola2018,wu2011who,lehmann2012dynamical}.

From the perspective of SLT, such markers can be interpreted as elementary
quanta of social excitation that raise the energy level of social atoms.
Repeated exposure to these signals through media systems, advertising
campaigns, or political messaging may gradually increase the excitation
level of a population, preparing conditions analogous to population
inversion in laser physics.

The amplification process itself depends critically on the emergence of
collective alignment within the informational field. In physical lasers,
phase coherence allows electromagnetic waves to interfere constructively.
In sociopolitical communication, a similar role may be played by the
alignment of interpretive frames, emotional orientations, and identity
markers across a population.

Section~7 of this article explores how such phase alignment relates to
well-established mechanisms in sociopolitical theory. These include
\textit{emotional contagion} \cite{Hatfield}, the
\textit{framing processes} through which social movements construct shared interpretations of events \cite{benford2000framing}, and the role of emotions in mobilizing collective action \cite{jasper2011emotions}. When communicative signals resonate with widely shared cognitive and affective structures, they can propagate rapidly through networks of social interaction.

Another important parallel concerns the statistical regimes of the field.
In physical lasers, photon statistics change as the system approaches the
lasing threshold: incoherent thermal radiation below threshold gives way
to partially coherent emission and eventually to a coherent state with
Poissonian photon statistics. Similar transitions may occur in the
dynamics of public discourse.

When informational signals are weakly correlated, communication resembles a noisy background of independent messages. As feedback mechanisms such as media amplification, social sharing, and algorithmic recommendation strengthen correlations between signals, collective attention may begin to synchronize. In some cases - such as viral advertising campaigns, political mobilizations, or mass protest movements - communication dynamics may approach a regime of coherent informational amplification.

These phenomena have been studied from several complementary perspectives in sociology and political science, including threshold models of collective behavior \cite{granovetter1978}, complex contagion theory
\cite{centola2018}, and research on collective attention in online
communication systems \cite{wu2011who,lehmann2012dynamical}. SLT does not replace these approaches but provides an additional conceptual framework that emphasizes the role of coherence, resonance, and stimulated informational emission in collective dynamics. Central to this framework is the representation of social atoms as complex information-processing agents inhabiting a high-dimensional state space. Unlike simpler models, the multi-marker approach treats the agent's internal configuration as a tensor product of marker-specific subspaces, where each marker corresponds to a distinct social topic or mental compartment. The interaction between these agents and the informational environment is modeled via a triple-resonance condition that integrates semantic overlap, energetic matching, and stance tolerance. By embedding these markers into a semantic manifold (or conceptual space \cite{SemM}), we account for associative priming and cognitive interference, where information about one topic can partially excite topologically adjacent markers. This interaction is formally described by an interaction Hamiltonian that couples the internal social energy of the atom to its stance, resulting in an asymmetric susceptibility to informational influence depending on the agent's current state. 

Despite these parallels, it is important to emphasize that both physical
and social lasers represent \textit{idealized theoretical constructs}.
Even in physics, coherent states and Poissonian photon statistics are
approximations to real systems affected by noise and environmental
fluctuations. Social systems are even more heterogeneous and subject to
continuous perturbations from competing narratives, institutional
structures, and external shocks.

Consequently, the analogies developed in SLT should be understood as
conceptual and mathematical modeling tools within the broader framework
of QLM rather than literal sociological descriptions.
Their value lies in providing a structured mathematical language for describing how aligned informational signals can generate rapid amplification of collective social behavior.

The application of SLT  is grounded in a broader foundational perspective that unifies quantum physics and QLM under the ``Bild'' (image) conception of scientific theories. 
As explored in Appendix, this approach moves beyond the ``straitjacket'' of identifying quantum observables with underlying ontic variables. Instead, it posits that both in the subatomic realm and in the macroscopic domains of cognition and sociology, the quantum formalism provides a blurred, 
approximate image - an echo of a complex reality. By treating the mathematical quantum structure as an epistemic tool rather than a literal 
ontological description, we can rigorously model the thresholds and leaps of social activation  while acknowledging that these processes 
emerge from a deterministic substrata of internal psychological states and social configurations that function as hidden variables within the QLM paradigm.

This paper is conceptual and aims to bridge SLT, and more generally QLM, with sociopolitical science. It is intended to attract the interest of researchers from both the natural sciences and the humanities. This bridging objective determines the multidisciplinary structure of the paper.

\section{Core Concepts of Social Laser Theory}

SLT offers a QL framework for modeling collective social behavior, particularly phenomena such as sudden mobilizations, protests, revolutions, or mass opinion shifts. Drawing inspiration from quantum optics - especially the principles underlying physical lasers - SLT conceptualizes society in terms of discrete agents (social atoms), quantized states of social energy, and coherent collective excitations driven by information fields. The theory is constructed as a mesoscopic model, bridging individual cognition and macro-social dynamics via an abstract formalism grounded in Hilbert space and information theory.

We provide an overview of the SLT methodology, focusing on core concepts rather than the underlying mathematical formalism or the quantum theory of lasers (for a detailed treatment, see \cite{laser1,laser2,laser5,laser6,laser12,laser13}). To assist the reader, each section is divided into two parts: a descriptive overview and a brief mathematical formalization. Those seeking a conceptual understanding may bypass the mathematical blocks during their initial reading.

\subsection{Social atoms as information-processing agents}

In a cognitively grounded interpretation, the social atom is modeled as an information-processing agent situated within a complex social environment. Rather than viewing it as a purely abstract unit, we treat it as a bounded rational cognitive system endowed with memory, attention, and motivational dispositions. To formalize this, we employ {\it Weberian ideal-typical markers}, allowing us to represent the agent's internal configuration as a tensor product of marker-specific subspaces. This approach acknowledges that while the agent's behavior is shaped by deep-seated representational structures and affective orientations, these markers remain theoretical constructs designed to map the leaps in social activation rather than literal, exhaustive psychological descriptions.

\subsubsection{Social energy}
One of the central dynamical variable of a social atom is its level of \emph{social energy} $E$. This energy is not physical but socio-cognitive: it reflects an individual's readiness for socially meaningful action. One may interpret $E$ as an integrated measure of motivational intensity, attentional engagement with socially relevant information, emotional arousal, and commitment to particular frames of interpretation. In this sense, social energy captures the degree to which an individual is prepared to move from passive observation to active participation.

A critical feature of this framework is that the quantization of energy levels is not merely a mathematical convenience, but a formalization of the ``thresholds and leaps'' observed in empirical social dynamics \cite{granovetter1978}. Empirically, the model recognizes that social movements do not grow at a constant, smooth rate; rather, they remain dormant until specific triggers cause a rapid transition, such as the leap from apathy to activism. Drawing on established models of social cascades, we recognize that individuals often remain in a state of stasis until a critical boundary of social influence or internal activation is crossed \cite{watts2002, schelling1971}.

Furthermore, psychological research into commitment suggests that internal activation is not a sliding scale but a series of qualitative states \cite{lewin1951}. These discontinuous transitions suggest that internal activation must reach specific, quantized levels before a qualitative change in behavior is triggered. Once an agent's motivational intensity reaches a specific threshold, a ``quantized'' shift occurs, transitioning the agent from passive observation to active mobilization. By applying the language of quantization to these psychological and sociological boundaries, we gain a rigorous method for representing the non-linear, ``all-or-nothing'' nature of human commitment \cite{centola2018}.

\subsubsection{Social atoms as two-level quantum systems}

The state of each social atom $i$ is modeled as a qubit. Mathematically, this is the normalized linear combination of two basic states: $|0\rangle_i$, representing a passive, uninformed, or disengaged agent, and $|1\rangle_i$, representing an excited, informed, or activated agent. In sociopolitical terms, the internal powers of the state and existing Foucauldian power relations provide the bias 
or pre-set of the social atom; they define the ``ground state'' ($|0\rangle$) of the population as a baseline of normalization. Transitions to the excited state $|1\rangle$ then represent a departure from this structural baseline toward active social mobilization.
The mathematical machinery is based on raising/lowering operators:
\[ \sigma_i^+ = \ket{1}_i \bra{0}, \quad \sigma_i^- = \ket{0}_i \bra{1}, \quad \sigma_i^z = \ket{1}_i \bra{1} - \ket{0}_i \bra{0} \]
satisfy the commutation relations: $[\sigma_i^+, \sigma_i^-] = \sigma_i^z, \quad (\sigma_i^+)^2 = (\sigma_i^-)^2 = 0 .$

In section \ref{State} we proceed with more complex model. 

\subsection{Quantum information field  as a structured information environment}

Social energy transitions are driven by interaction with the \emph{infon field} - a socially shared information field composed of discrete excitations called \emph{infons}. Analogous to photons in quantum electrodynamics, infons carry quantized units of ideological, emotional, or symbolic content. They are generated and circulated by media, political actors, or discourse networks, and their repeated interaction with social atoms can induce state transitions.

Each infon is characterized by its energy content, frequency, and mental markers. In this framework, we distinguish between cognitive markers - related to thinking, reasoning, and information processing - and affective markers - related to emotions, feelings, and mood. Due to the highly interconnected and fuzzy nature of human psychology, we treat these concepts as Weberian ``ideal types'' (mental images rather than absolute entities).By acknowledging that these are theoretical images (\textit{Bild}) - or "Notions" with a Platonic capital N - we purposefully intensify certain features to match abstract terminology while diminishing others. This allows us to project the multidimensionality of human thought into a clear, dichotomous state space ($m = \pm$). While actual mental representations are "put into the mold" of these ideal-typical pictures - involving a degree of amplification and contraction of features - this idealization is essential for the mathematical formalization of the model. It enables the identification of the specific thresholds and leaps necessary for the triple-resonance mechanism developed in Section \ref{triple}, where an infon must match the agent's internal state through a simultaneous alignment of energy, frequency, and idealized mental markers \cite{Weber, Plato, Aspalter1, Aspalter2}.

The use of Weberian 'ideal types' here is not merely metaphorical; it serves as a crucial methodological bridge. In sociopolitical science, an ideal type is a simplified, intensified model used to categorize complex reality into clear, analytical constructs. By treating cognitive and affective markers as ideal types, we perform a move similar to the 'idealized modeling' found in physics (such as the point-particle or ideal gas). This allows us to project the 'fuzzy' and continuous nature of human psychology into the discrete, dichotomous state space ($|0\rangle$ and $|1\rangle$) required for quantum-like modeling, providing the necessary theoretical yardstick to measure social resonance.

To be sure, if e.g. the affective marker component of a mental marker is essentially stronger than its cognitive component, we can treat it, in the ideal-typical Weberian sense, as an affective marker. This of course also applies the other way around.

This is needed and functions aptly in the research set-up pursued here and goals envisioned. When it comes to measurements the set-up then, as follows, allows for superposition states, where may measure a `full yes' and a `full no' at the same time, at once. Inner products between measured states represent the degrees of overlap in between different abstract entities, dimensions and variables (e.g. cognitive versus affective markers), as every such abstract concept must also always allow for continuous dimension of degrees of meaning, also enabling mixed measurements of pure abstract meanings (mixed states) and meaning splits (superpositions).

\subsection{Mental - cognitive and affective - markers}

So, each infon carries a mental marker $m$ that encodes aspects of its conceptual meaning, identity relevance, and emotional tone. At the same time, each agent possesses an internal configuration of markers shaped by memory, social identity, ideological commitments, and affective history. When there is strong alignment between the marker structure of an infon and the internal configuration of the agent, i.e. with a high degree of parallelity, the probability of absorption increases significantly.

This alignment can be formalized as similarity in a multidimensional semantic-affective space. In more intuitive terms, resonance occurs when a message ``fits'' how individuals already think and feel. Under conditions in which many agents share similar cognitive frames and affective orientations, the same infons will resonate broadly. The result is synchronized activation across a population. In sociological language, this corresponds to processes of frame alignment, emotional contagion, and identity-based mobilization. Within the Hilbert space formalism of SLT, such large-scale synchronization is represented as a coherent state $|\alpha\rangle$, but substantively it describes the collective alignment of attention, interpretation, and motivational readiness.

Each mental marker $m$ has two components, $m=(c,a),$ where $c$ and $a$ are cognitive and affective markers.   
A {\it cognitive marker} $c$ is a feature or symbolic element that structures reasoning and interpretation. It may take the form of a concept, belief, narrative template, causal model, or interpretive frame.
An {\it affective marker} $a,$ by contrast, encodes emotional valence and arousal. It shapes motivation, attention, and memory consolidation by attaching feeling tones to representations. 

Importantly, cognitive and affective markers are deeply interconnected. Emotions influence how evidence is interpreted, while interpretive frames structure emotional responses. In SLT this interconnection is mathematically described as {\it entanglement of cognitive and affective markers}, see article \cite{marker} for details and examples of such entanglement. 

\section{Quantum state spaces of social atoms and infons}  

\subsection{Quantum representation of the multi-marker social atom}
\label{State}

 This framework models a social atom (social agent) as a composite quantum system characterized by $N$ distinct mental markers. These markers represent the internal degrees of freedom of the social atom and interact dynamically with a quantized information field consisting of discrete units of influence, termed infons. 

\paragraph{Marker value and dimensionality.} For the sake of analytical clarity, we consider dichotomous markers ($m = \pm$). In this binary configuration: $m = +$ represents a positive or supportive attitude toward a specific social object (e.g., the government); $m = -$ represents a negative or oppositional attitude.

While the dichotomous model serves as the foundational ``spin-1/2'' analogue for social behavior, the framework is naturally extensible to multi-valued markers. For instance, a three-level system can be employed to account for a neutral stance: $m = \{-, 0, +\}$ where $m = 0$ explicitly quantifies a state of neutrality or informational indifference. In the general case, each marker $j$ (where $j = 1, \dots, N$) inhabits a local Hilbert space $\mathcal{H}_j$ whose dimension is determined by the number of possible attitudes associated with that specific marker which is multiplied by the number of social energy levels characterizing a marker. Generally each marker has its own energy spectrum $E_{0, m} < ...< E{n_m, m}.$ Again for simplicity we consider the model in that all markers have two energy levels.  

\paragraph{Social atom's state space.} The Hilbert space of the social atom operating with $N$ dichotomous mental markers having two level social energy spectra is  the tensor product of $N$ marker-specific subspaces:
\begin{equation}
\mathcal{H}_{atom} = \bigotimes_{j=1}^N \mathcal{H}_j, \quad \text{dim}(\mathcal{H}_{atom}) = 4^N
\end{equation}
Each subspace $\mathcal{H}_j$ corresponds to the $j$-th mental marker and is spanned by four basis states representing the combination of social energy (passive/active) and marker value (positive/negative):
\begin{equation}
\mathcal{B}_j = \{ |E_{0j}, +\rangle, |E_{0j}, -\rangle, |E_{1j}, +\rangle, |E_{1j}, -\rangle \}
\end{equation}

\paragraph{Hamiltonian  of a social atom ($\hat{H}_{atom}$).}

The internal energy of the atom is the sum of the energies of all its marker constituents. The energy required for a social transition in the $j$-th marker is the difference between the excited and ground levels: $\Delta E_j = E_{1j} - E_{0j}$.
\begin{equation}
\hat{H}_{atom} = \sum_{j=1}^N \sum_{m \in \{+,-\}} \left( E_{1j} |E_{1j}, m\rangle\langle E_{1j}, m| + E_{0j} |E_{0j}, m\rangle\langle E_{0j}, m| \right)
\end{equation}
Using the operator $\hat{\sigma}_z^{(j,m)} = |E_{1j}, m\rangle\langle E_{1j}, m| - |E_{0j}, m\rangle\langle E_{0j}, m|$, this is expressed as:
\begin{equation}
\hat{H}_{atom} = \sum_{j=1}^k \sum_{m \in \{+,-\}} \frac{\Delta E_{j}}{2} \hat{\sigma}_z^{(j,m)} + \text{const.}
\end{equation}

\subsection{Quantum information field formalism}

This formalism describes the interaction between a social agent and a quantized information field, where energy, topic (marker), and stance (value) are treated as independent degrees of freedom.

\paragraph{The single-infon Hilbert space ($\mathcal{H}_{infon}.$)}
An infon is a quantum of information defined by three independent variables. Its Hilbert space is the tensor product:
\begin{equation}
\mathcal{H}_{infon} = \mathcal{H}_E \otimes \mathcal{H}_{marker} \otimes \mathcal{H}_{value}
\end{equation}
A basis state $|E, j, v\rangle$ represents an infon with:
\begin{itemize}
    \item $E \in \mathbb{R}^+$: The \textbf{Energy} (impact/weight of the information).
    \item $j \in \{1, \dots, N\}$: The \textbf{Marker Index} (the specific social topic).
    \item $v \in \{+, -\}$: The \textbf{Marker Value} (the stance or attitude).
\end{itemize}
These states satisfy $\langle E', j', v' | E, j, v \rangle = \delta(E - E') \delta_{jj'} \delta_{vv'}$.

\paragraph{The marker Hilbert space: $\mathcal{H}_{marker}.$}

The marker $\mathcal{H}_{marker}$ is $N$-dimensional Hilbert space that defines the ``address'' in the form of a mental marker of an infon. It represents the $N$ distinct markers or mental compartments that a social agent can possess.

This is the good point to make the following foundational remark. The marker decomposition of a quantum information field depends not only on its source (mass-media, social networks,...), but also on the population of social atoms exposed by the ``radiation'' of this field. It can happen that two different populations decompose the same information field into a different collection of mental markers.  

The marker space is isomorphic to $\mathbb{C}^N$: $\mathcal{H}_{marker} \cong \mathbb{C}^N.$
This space is spanned by an orthonormal basis of the eigenstates of the marker operator,
 $\mathcal{B}_{marker} = \{ |j\rangle \}_{j=1}^N
$
where each $|j\rangle$ corresponds to a unique mental marker.
The basis states satisfy the standard Kronecker delta orthonormality condition:
$\langle j' | j \rangle = \delta_{j'j}.$
This is the important constraint of the model; more generally we can proceed even with non-orthogonal mental markers. 

\paragraph{Social interpretation.}
In the social context, $\mathcal{H}_{marker}$ represents the categorical taxonomy of the agent's interaction with the information field.
 An infon in a state $|j\rangle$ is ``unambiguous'' information regarding a single topic (including its emotional coloring).
An infon can exist in a superposition $|\phi\rangle = \sum c_j |j\rangle$, representing ``interdisciplinary'' or ``complex'' information that addresses multiple mental markers simultaneously.

\textbf{Information Routing:} Without this space, information would be a global field affecting all mental aspects equally. $\mathcal{H}_{marker}$ allows for the localization of informational impact.

\paragraph{The Fock space ($\mathcal{H}_{field}$).}
The information environment is a bosonic Fock space constructed from $\mathcal{H}_{infon}$, allowing for an indefinite number of infons $n$:
\begin{equation}
\mathcal{H}_{field} = \bigoplus_{n=0}^{\infty} \text{Sym}(\mathcal{H}_{infon}^{\otimes n}).
\end{equation}
The field is governed by operators $\hat{a}^\dagger(E, j, v)$ and $\hat{a}(E, j, v)$ satisfying:
\begin{equation}
[\hat{a}(E, j, v), \hat{a}^\dagger(E', j', v')] = \delta(E - E') \delta_{jj'} \delta_{vv'}
\end{equation}
We mention some special states.  The social vacuum state $|0\rangle$ (here $n=0)$ represents a silent environment devoid of disseminated information. It is the unique state such that $\hat{a}(E, j, v)] |0\rangle=0$ for all $(E, j, v).$
A single-infon state $\hat{a}^\dagger(E, j, v)|0\rangle$  represents the presence of exactly one unit of information (e.g., a single message) carrying the concrete mental marker endowed with some amount of social energy. High-intensity states 
  $\hat{a}^\dagger(E_n, j_n, v_n) \cdots \hat{a}^\dagger(E_1, j_1, v_1)|0\rangle$
(large values of $n$) represent a dense informational atmosphere (e.g., a viral social media campaign or mass propaganda).

The total energy of the information environment (the Field Hamiltonian) is:
\begin{equation}
\hat{H}_{field} = \sum_{j=1}^N \sum_{v \in \{+,-\}} \int E \, \hat{a}^\dagger(E, j, v) \hat{a}(E, j, v) \, dE
\end{equation}

\section{Quantum Dynamics of the Triple-Resonant Social Atom}
\label{triple}

Large-scale social mobilization exhibits structural similarities to a laser's transition to coherent emission. This process is governed by the Interaction Hamiltonian ($\hat{H}_{int}$), which is sensitive to the agent's current state. Sociopolitically, this sensitivity is shaped by Foucauldian power relations and the internal powers of the state, which create a background atmospheric interpellation. In our model, these power dynamics are captured by the Lorentzian energy-coupling function ($g_{k,v}(E)$), which determines the trigger point for social activation based on an individual's position within these pre-existing force fields.

The interaction between a social atom and the information field is governed by three simultaneous resonance conditions: 
semantic (marker), energetic, and valence (stance) resonance. Crucially, the internal social energy of the atom is coupled to its stance, creating an asymmetric susceptibility to information.

\paragraph{Social energy coupling ($g_{k,v}(E)$).}
The energy matching is enforced by a Lorentzian profile. The transition energy $\Delta E_{k,v} = E_{1,k,v} - E_{0,k,v}$ depends on both the topic $k$ and the stance $v \in \{+,-\}$. This allows for modeling agents who are more easily "triggered" or "convinced" in one stance over another:
\begin{equation}
g_{k,v}(E) = \frac{V_0 \gamma}{(E - \Delta E_{k,v})^2 + \gamma^2}
\end{equation}

\paragraph{Semantic overlap ($\Gamma_{kj}$) and topological mapping.} 
To move from the discrete description of  markers to continuous (which is convenient for modeling of resonance), we  embed both the mental markers of the social atom and the infon field into a common semantic manifold $\mathcal{M}_{sem}$ (cf. conceptual space
\cite{SemM}).
 Each marker $k$ is characterized by a central coordinate $x_k \in \mathcal{M}_{sem}$, representing its core thematic ``anchor,'' while each infon $j$ carries a coordinate $x_j$. 

The overlap coefficient $\Gamma_{kj}$ is defined as a spatial resonance function, typically modeled as a Gaussian kernel:
\begin{equation}
\Gamma_{kj} = \exp\left( -\frac{d(x_k, x_j)^2}{2\sigma^2} \right)
\end{equation}
where $d(x_k, x_j)$ is the geodesic distance on the semantic manifold (e.g., Euclidean distance in a simplified linear spectrum) and $\sigma$ represents the thematic width or ``cognitive aperture'' of the marker. 
This formulation implies that a social atom does not perceive information in isolation. Instead, an infon of topic $j$ generates a \textit{distribution of excitation} across all markers $k$ that are topologically adjacent. This accounts for 
 associative priming: information about one topic partially excites related markers due to the non-zero tails of 
the Gaussian resonance, and for marker interference: f two markers $k_1$ and $k_2$ are semantically close ($|x_{k1} - x_{k2}| < \sigma$), they may compete for the same infon energy, leading to cognitive dissonance or reinforcement.

\paragraph{Stance Resonance ($\Lambda_{v,v'}$).}
As for mental markers, it is convenient to proceed with  fuzzy stances described by a continuous stance set, e.g. $v \in [0,1].$  We define a stance overlap between the marker's internal value $v$ and the infon's stance $v'$. Using a discrete or continuous representation:
\begin{equation}
\Lambda_{v, v'} = \exp\left( -\frac{(v - v')^2}{2\eta^2} \right)
\end{equation}
where $\eta$ is the tolerance width. This ensures that the interaction is suppressed if the infon's stance is too far from the marker's current state.

\paragraph{The Interaction Hamiltonian.}
The Hamiltonian incorporates the triple-matching condition, where the energy resonance $g_{k,v}(E)$ is now sensitive to the atomic stance:
\begin{equation}
\hat{H}_{int} = \sum_{k,j} \sum_{v, v'} \int \Gamma_{kj} \Lambda_{v, v'} g_{k,v}(E) \left( \hat{\sigma}_+^{(k, v)} \hat{a}(E, j, v') + \hat{\sigma}_-^{(k, v)} \hat{a}^\dagger(E, j, v') \right) dE
\end{equation}

\paragraph{Derivation of Transition Probability $P_{0 \to 1}$.}
To find the probability of a marker $k$ with stance $v$ transitioning due to an infon $(E, j, v')$, we solve for $c_1(t)$ in the interaction picture:
\begin{equation}
i\hbar \frac{d}{dt} c_1(t) = \Gamma_{kj} \Lambda_{v, v'} g_{k,v}(E) e^{i(\Delta E_{k,v} - E)t/\hbar}
\end{equation}
Integrating from $0$ to $t$ yields the transition probability:
\begin{equation}
P_{0 \to 1}(t, E, j, v, v') = |\Gamma_{kj}|^2 |\Lambda_{v, v'}|^2 |g_{k,v}(E)|^2 \frac{4 \sin^2\left( \frac{(\Delta E_{k,v} - E)t}{2\hbar} \right)}{(\Delta E_{k,v} - E)^2}
\end{equation}

\paragraph{Determinants of Infon–Marker Coupling.}

The transition probability was highly affected the coupling between the infon in the quantum information field and the corresponding marker in the social atom, including the social energy couping, semantic overpay, and stand resonance.

First, energy refers to the impact or weight of the information. In social learning \cite{1.}, people tend to learn from individuals to whom others have preferentially attended, learned or deferred, i.e., prestige bias  \cite{2.}. This tendency was further amplified by the algorithms of relevant platforms  \cite{3.}.

Second, the semantic overlap between infon in the quantum information field and that of the social agent also has an significant impact on the coupling. To empirically explain how the semantic overlap can have a substantial impact on the effect of an incoming infon on the transition probability, we can compare how the same infon differs in its effect among different quantum information fields. Previous literature  \cite{1.} has found that participants tend to disproportionately learn from their ingroups  \cite{4.}. In digital world, the ingroup information were then amplified  \cite{5.} by the platform algorithms. The two processes working together make the users to produce more ingroup information  \cite{6.}. Two important groups among others are empirically studied in literature. The first one is the cultural group  \cite{7.}. Traditionally, the world was culturally divided into individualistic and collectivistic societies. In the former society, the self was portrayed as an independent entity composed of internal attributes such as personality traits, motives, goals, and attitudes. In the latter society, the self was described as interdependent with others in their ingroups is prevalent. The interdependent self is characterized centrally by relational attributes, such as social roles and status, which defines one’s identity. As a result of the cultural difference, peoples in collective society tend to have a holistic cognitive style and to have more intuitive or less categorical modes of inference  \cite{8.}. The second one is the language group. It is widely believed that how persons think, perceive and understand the world is highly influenced by the language they speak, i.e. the Spair-Whorf hypothesis  \cite{9.}. Previous literature have found that language can play a significant role in structuring, or restructuring, a domain as fundamental as spatial cognition  \cite{10.} and color perception  \cite{11.}.

Third, the moral and emotional value can also have a significant effect on the coupling of the infon and the social agent. Previous literature  \cite{1.} have found that people tend to pay outsized attention to moralized  \cite{12.} and to emotionally arousing information, specically negatively valenced information, especially when such information is relatively rare  \cite{13.}. In digital platforms, these moralized and emotional information were then amplified by the algorithms  \cite{14.} of the platforms, which makes users in the platforms to produce more moralized and emotional  \cite{15.} information.

\paragraph{The multi-Key selection mechanism.}
The atom's internal state is modified only when the information field satisfies a triple-key resonance:
\begin{enumerate}
    \item \textbf{Topic Relevance:} The infon amrker $x_j$ must be semantically close to social atom's marker $x_k$.
    \item \textbf{Stance Tolerance:} The infon stance $v'$ must match the marker stance $v$ within the width $\eta$.
    \item \textbf{Specific Energetic Resonance:} The infon energy $E$ must match the gap $\Delta E_{k,v}$. Importantly, if $\Delta E_{k,+} \neq \Delta E_{k,-}$, the agent requires different levels of informational impact to change their mind depending on their current stance.
\end{enumerate}
To simplify the model we proceeded with mental markers as integral labels that is without splitting them into cognitive and affective components. The model can be generalized to cognitive-affective markers framework. 

\section{Affective markers: from social mobilization and viral advertising to social lasing}

In political campaigns (especially leading to mass protests), as well as in advertising strategies aimed at generating new needs  -  such as emerging consumer trends or shifts in lifestyle  -  communicators often rely primarily on affective techniques that carry minimal semantic load. These techniques make it possible to introduce an idea into mass discourse in the most cost-effective and wide-reaching manner. The central requirement for affective markers is singular: they must evoke strong emotions. The specific type of emotion is of little importance - anger and disgust are just as effective as amusement or laughter.

From a theoretical perspective, such affective markers can be interpreted as semiotic devices functioning predominantly on the level of connotation rather than denotation, see Barthes \cite{Barthes1972}. Their communicative power resides not in propositional content but in their ability to mobilize emotional resonance, thereby shaping the field of possible meanings available to collective interpretation. Within discourse analysis, their circulation exemplifies what Laclau \cite{Laclau2005} describes as the process of affective investment in signifiers, through which empty or floating signifiers acquire mobilizing force by becoming saturated with emotion. This mechanism also resonates with Habermas's \cite{Habermas1984} concern that strategic communication, when divorced from rational argumentation, risks privileging manipulation over deliberation. In the domain of communication theory, such techniques align with models of low-information, high-impact messaging \cite{McCombs,Entman}, where affective intensity, rather than informational depth, drives agenda-setting, framing, and the saturation of discourse.

Empirical studies on viral advertising confirm and operationalize these theoretical insights. Eckler and Bolls \cite{Eckler} show that the emotional tone of viral messages significantly influences both attitudes toward the advertisement and intentions to forward it, highlighting the role of affect as a transmission amplifier in digital networks. Dafonte-Gomez \cite{Dafonte} further identifies the core elements that make viral content successful, emphasizing that motivational and emotional triggers - rather than rational appeals - are the primary determinants of sharing behavior. More recently, Rachmad \cite{Rachmad1,Rachmad2} in two complementary reports for the United Nations Economic and Social Council, explores the dual role of ``viral'' and ``gimmick'' effects in shaping behavioral transformation. His findings suggest that gimmicks, understood as low-information but high-salience communicative acts, can function as catalysts for behavioral contagion and discourse alignment. Together, these studies provide empirical grounding for the notion that affective markers operate as minimal yet potent carriers of influence in contemporary communication ecosystems.

When viewed through the lens of SLT, affective markers  may be conceptualized as discrete ``quanta'' of social energy that resonate within a communicative medium. Much like photons in physical lasers, these emotionally charged semiotic units are capable of exciting social atoms (individuals or groups) into higher levels of social energy (typically affective arousal). The repetition and amplification of such markers within media discourse create conditions of coherence, whereby emotional responses become synchronized across large populations. In this respect, the affective markers identifiedare  the minimal yet potent informational carriers that trigger resonance, phase alignment, and ultimately large-scale collective mobilization. Thus, classical theories of framing and agenda-setting can be recast within SLT as mechanisms that structure (align and amplify) the energy landscape of discourse, determining which emotional excitations reach threshold levels necessary for social coherence and action.

\section{Coherent States in Physical Lasers}

\subsection*{Coherent States of the Electromagnetic Field}

In quantum optics, a coherent state \( |\alpha\rangle \), where $\alpha$ is a complex number,  is a minimum uncertainty state of the quantized electromagnetic field, defined as the eigenstate of the annihilation operator:
\[
a|\alpha\rangle = \alpha |\alpha\rangle.
\]
Its Fock state expansion is:
\[
|\alpha\rangle = e^{-|\alpha|^2/2} \sum_{n=0}^\infty \frac{\alpha^n}{\sqrt{n!}} |n\rangle,
\]
where \( |n\rangle \) denotes the photon number state with \( n \) photons.

These states exhibit classical-like behavior with well-defined amplitude and phase, and a Poissonian photon number distribution:
\begin{equation}
\label{Poisson}
P(n) = \frac{|\alpha|^{2n} e^{-|\alpha|^2}}{n!}.
\end{equation}

\subsubsection*{Role in Lasers}

In a laser operating above threshold:
\begin{itemize}[noitemsep]
    \item Stimulated emission leads to exponential amplification of photons in a single cavity mode.
    \item The emitted photons are in phase with the preexisting field, resulting in phase-coherent light.
    \item The quantum state of the field is well approximated by a coherent state \( |\alpha\rangle \), especially when the average photon number \( \langle n \rangle = |\alpha|^2 \gg 1 \).
\end{itemize}

We highlight that the real field in the laser cavity is not exactly a pure coherent state due to quantum noise and decoherence, but it is very well approximated by a statistical mixture centered around \( |\alpha\rangle \):
\[
\rho = \int P(\alpha) |\alpha\rangle\langle\alpha|\, d^2\alpha,
\]
where \( P(\alpha) \) is sharply peaked for a stable laser mode.

\section{Coherent States and Activation Dynamics in Social Lasers}

\subsubsection*{Coherent Infon Fields}

In SLT, the quantum information field - \emph{infon field} - plays the role of the electromagnetic field in a physical laser. Each infon is a quantum of structured, stimulating social information. Each mode of the inforn field is characterized by its social energy $E$ 
 and a set of mental markers $m=(m_1,...,m_s)$ which may correspond to a specific 
discourse frame (e.g., anti-authoritarianism, nationalism, etc.).    

A coherent state of infons for mode \( k=(E,m) \), 
\begin{equation}
\label{cstate}
|\alpha_k\rangle = e^{-|\alpha_k|^2/2} \sum_{n=0}^{\infty} \frac{\alpha_k^n}{\sqrt{n!}} |n\rangle_k,
\end{equation}
satisfies the eigenvalue equation $b_k |\alpha_k\rangle = \alpha_k |\alpha_k\rangle.$
Here, \( b_k \) is the annihilation operator for infon mode \( k \).

These coherent infon states:
\begin{itemize}[noitemsep]
    \item Encode highly ordered, low-entropy semantic configurations,
    \item Support amplification through phase-aligned social activation,
    \item Model viral or resonant ideologies, e.g., slogans or narratives spreading coherently across a population.
\end{itemize}

\subsubsection*{Population Inversion in Society}

The social analog of atomic population inversion is a population of agents predominantly in the activated state:
\[
|\Psi_{\text{soc}}\rangle = |1\rangle^{\otimes N}.
\]
This state represents a society ready for collective action or resonance with a dominant ideological signal.

Once a coherent infon probe (e.g., slogan, meme) interacts with this inverted - i.e. excited and receptive - population, it triggers \textit{stimulated activation}, agents emit new infons aligned with the dominant mode, reinforcing the signal. In SLT the infon field state can be approximately described as  |$\alpha_k\rangle.$  

\subsection*{Interpretation}

Coherent infon states can be understood as modeling structured ideological discourse fields in which meanings are not randomly distributed but organized in a phase-aligned and mutually reinforcing way. Here, the parallelity of aggregate infon states, or infon fields, is translated into collective meaning-creation that give rise to socio-political force fields: such as the construction and amplification of ideological fields, political and socio-cultural atmospheres, increasing polarization of asymmetric power relations, the widening emotional/political gap between rulers (the super-rich and the powerful) and the common people, and so forth.

In this setting, population inversion corresponds to a condition of social readiness for mass mobilization: a situation in which a sufficiently large portion of the population occupies an excited cognitive or emotional state, making large-scale activation possible. The presence of coherence is crucial, because it enables the exponential amplification of meaning, allowing certain narratives or symbols to grow rapidly in intensity and influence.

Within the SLT framework, the introduction of coherent states provides a powerful conceptual tool for explaining how collective resonance can emerge from communication that would otherwise appear noisy, fragmented, or weakly structured.

In other words, in relational/conformal social quantum field theory terms, angular orientations have been polarized into pro this and contra that, each group/pole in opposing terms to the other. Within the prepared/excited (or activated) population, also called inverse population in laser theory, angular orientations have been changed, relationships altered, meanings, motivations and readiness to act have been amplified and/or diametrically altered (pluses turned into minuses and vice versa). In general, relative degrees of similarities alter the course of politics, economics and the society as a whole \cite{Aspalter5,Aspalter6, Aspalter7}.

\section{The Phase of a Coherent State in Physical and Social Lasers}

In the definition of a coherent state \( \alpha \in \mathbb{C} \) is a complex number. This number encodes both the amplitude and phase of the state $\alpha = |\alpha| e^{i\theta},$
with:
\begin{itemize}
    \item \( |\alpha| \): the amplitude (related to the average photon number),
    \item \( \theta \): the phase of the coherent state.
\end{itemize}

In quantum optics, the phase \( \theta \) determines the location of the coherent state in phase space. The expectation value of the field operator is $\langle \alpha | b | \alpha \rangle = \alpha = |\alpha| e^{i\theta}.$
Using the quadrature operators:
\[
q = \frac{1}{\sqrt{2}}(b + b^\dagger), \quad p = \frac{1}{\sqrt{2}i}(a - a^\dagger),
\]
the coherent state's center in phase space is:
\[
\langle q \rangle = \sqrt{2} |\alpha| \cos\theta, \quad \langle p \rangle = \sqrt{2} |\alpha| \sin\theta.
\]

\begin{center}
\begin{tikzpicture}[scale=1.4]
  \draw[->] (-2,0) -- (2,0) node[right] {$q$};
  \draw[->] (0,-2) -- (0,2) node[above] {$p$};
  \draw[->, thick, blue] (0,0) -- (1.4,1.4) node[anchor=west] {$\alpha = |\alpha| e^{i\theta}$};
  \draw[dashed] (1.4,0) -- (1.4,1.4) -- (0,1.4);
  \node at (1.5,-0.2) {$\sqrt{2}|\alpha|\cos\theta$};
  \node at (-0.3,1.4) {$\sqrt{2}|\alpha|\sin\theta$};
  \draw[->] (0.5,0) arc (0:45:0.5);
  \node at (0.7,0.2) {$\theta$};
\end{tikzpicture}

\vspace{0.5em}
Figure: Coherent state \( |\alpha\rangle \) as a point in phase space.
\end{center}

In laser physics, the phase \( \theta \) plays a central role in determining how the electromagnetic field behaves under interference. When several coherent beams overlap, it is their relative phases that decide the outcome: aligned phases lead to constructive interference and amplification, while opposite phases produce destructive interference and attenuation. A well-functioning laser emits light that is coherent and characterized by a well-defined phase, although, on longer time scales, unavoidable quantum fluctuations can gradually induce phase diffusion.

The number of quanta (photons) in the coherent state \( |\alpha\rangle \) follows a Poisson distribution (\ref{Poisson}). This statistical distribution depends only on the magnitude \( |\alpha| \), which sets the average photon number, and is independent of the phase \( \theta \). Nevertheless, while the phase does not influence the photon-number statistics directly, it remains essential for interference and coherence phenomena, where relative phase differences determine observable effects.

In SLT, coherent states of the infon field are used to model structured ideological fields. The phase of a coherent state
$|\alpha_k\rangle$ can be interpreted as follows. The phase \( \theta_k \) represents the alignment of emitted infons with the dominant ideological or political direction. If social agents emit infons with the same phase \( \theta_k \), their communication is constructive and reinforcing. Misaligned phases lead to destructive interference, noise, or fragmentation of discourse.

\subsubsection*{Collective Coherence}

The phase coherence of social infons is what makes large-scale amplification possible in the first place. When individual cognitive excitations share a common phase, ideological signals can be reinforced through a mechanism analogous to stimulated emission, so that meanings do not merely circulate but grow in intensity. This shared phase alignment also enables synchronization of collective behavior: individuals respond in a coordinated way, leading to coherent activation rather than scattered, uncorrelated reactions. Under such conditions, certain narratives, symbols, or slogans can rapidly emerge as dominant, since their repetition occurs in phase across the population and thus accumulates constructively.

\textbf{Summary.} The phase \( \theta \) of a coherent state plays a fundamental role both in quantum optics and in quantum-like models of society. Although it does not influence number statistics directly, it governs coherence, interference, and alignment within the field. In SLT, coherence of phases across the population is a necessary condition for ideological amplification and collective resonance. In this sense, SLT mirrors the functioning of physical lasers, but transposes these mechanisms into a cognitive and discursive space, where alignment of meaning replaces alignment of electromagnetic waves. In the following, the cognitive and discursive space results in the formation of mass social action space.

\section{Photon Statistics in Physical Lasers}

The way photons are distributed inside a laser cavity depends strongly on how the laser is operating. In fact, the statistical properties of the light change qualitatively as the system moves from weak excitation to fully developed lasing.

When the laser operates \emph{below threshold}, emission is dominated by spontaneous processes. The light behaves essentially like thermal radiation, and the photon number follows a Bose--Einstein distribution. In this regime fluctuations are large: the variance exceeds the mean, $\mathrm{Var}(n) > \mathbb{E}[n]$. Physically, this means photons tend to arrive in bunches, and the field lacks stable phase coherence.

As the system approaches the \emph{threshold}, it enters a transitional regime. Stimulated emission becomes increasingly important, but spontaneous emission and noise are still significant. The photon statistics are no longer purely thermal, yet not fully coherent either. Fluctuations are often enhanced due to strong feedback in the cavity, and the system displays critical-like behavior.

In the idealized picture of a laser operating \emph{above threshold}, the intracavity field approaches a coherent state $|\alpha\rangle$, defined by
\[
a|\alpha\rangle = \alpha|\alpha\rangle.
\]
In this state the photon number follows a Poisson distribution,
\[
P(n) = \frac{|\alpha|^{2n}}{n!} e^{-|\alpha|^2},
\]
with mean photon number $\bar{n} = |\alpha|^2$ and variance $\mathrm{Var}(n) = \bar{n}$. The equality of mean and variance is the statistical signature of ideal coherent light: fluctuations are present, but they are minimal and well-balanced.

Real lasers, however, never perfectly realize this ideal limit. Even above threshold, several physical mechanisms introduce deviations from exact Poisson statistics. Phase diffusion broadens the spectral line, pump noise injects additional randomness, gain saturation modifies amplification dynamics, and quantum fluctuations or mode competition further perturb the field. Depending on conditions, these effects can produce either sub-Poissonian statistics, where $\mathrm{Var}(n) < \bar{n}$, or super-Poissonian statistics, where $\mathrm{Var}(n) > \bar{n}$. Thus, realistic lasers display a rich variety of fluctuation behaviors.

\subsection*{Second-order Coherence and Fluctuation Measures}

Photon statistics are often characterized through the second-order correlation function at zero delay,
\[
g^{(2)}(0) = \frac{\langle n(n - 1) \rangle}{\langle n \rangle^2},
\]
which quantifies the tendency of photons to bunch together. For thermal light one finds $g^{(2)}(0) = 2$, indicating strong bunching. For coherent light, $g^{(2)}(0) = 1$, reflecting Poissonian statistics and statistically independent emission events. Values $g^{(2)}(0) < 1$ correspond to antibunching, a genuinely quantum effect that cannot be explained classically.

Another useful quantity is the Fano factor,
\[
F = \frac{\mathrm{Var}(n)}{\mathbb{E}[n]},
\]
which compares the observed variance to that of a Poisson distribution. When $F \approx 1$, the field is close to coherent. If $F > 1$, fluctuations are enhanced, as in thermal or burst-like emission. When $F < 1$, fluctuations are suppressed, signaling nonclassical or squeezed light.

\section*{Analogy to Social Laser Theory}

In SLT, \emph{infons} - units of information carrying cognitive meaning - play a role analogous to photons. Extending the statistical analogy helps clarify different regimes of social information dynamics.

In a social system operating below threshold, individuals produce and share information in a largely uncoordinated way. Emission of infons is sparse, weakly correlated, and driven mostly by spontaneous personal motivations. The overall pattern resembles thermal radiation: noisy, scattered, and lacking collective alignment.

Near the social threshold, feedback mechanisms such as likes, shares, algorithmic amplification, or media reinforcement begin to synchronize agents. Infons carrying similar cognitive markers become increasingly correlated. Fluctuations grow, and bursts of coordinated communication start to appear. The system hovers between disorder and coherence.

In an idealized above-threshold social regime, a macroscopically coherent state emerges. Large groups of agents emit aligned infons, reinforcing the same narratives, slogans, or symbolic frames. Communication becomes strongly amplified and appears highly coordinated. Statistically, this resembles the Poissonian regime of a physical laser: emission is intense yet structurally ordered.

Real social systems, however, rarely maintain perfect coherence. Differences in attention, competing discourses, external shocks, and media noise introduce irregularities. As a result, communication patterns often exhibit bursts, cascades, and viral spikes - forms of super-Poissonian behavior in which fluctuations exceed the mean. In this way, the analogy with laser physics not only illuminates the possibility of coherent amplification, but also explains why real social dynamics remain intrinsically noisy and variable.

\begin{center}
\begin{tabular}{|c|c|c|}
 \hline
 \textbf{Regime} & \textbf{Physical Laser} & \textbf{Social Laser Analogy} \\
 \hline
 Below Threshold &
 Thermal (Bose--Einstein) &
 Weak, random infon emissions \\
 \hline
 Near Threshold &
 Transition regime &
 Feedback, mild coherence \\
 \hline
 Above Threshold (ideal) &
 Coherent state (Poisson) &
 Aligned infon emission (early coherence) \\
 \hline
 Above Threshold (real) &
 Deviates from Poisson &
 Bursts, cascades, viral spreading \\
 \hline
\end{tabular}
\end{center}

\section{Amplification through phase-aligned social sctivation: Analogies from sognitive and sociopolitical sciences}
\label{amp}

As discussed in the introduction, one of the central aims of this article is to develop a shared vocabulary that bridges the foundational concepts of SLT with corresponding notions in sociopolitical science. In this section, we explore the relationship between the SLT concept of \emph{amplification through phase-aligned social activation}---a verbal formulation of the coherent-state dynamics characteristic of the idealized social lasing regime---and a range of established ideas from sociopolitical theory, mainly relayed to 
resonance, cascade amplification, social coherence.  

We find that many key constructs in political psychology, communication studies, and identity theory can be meaningfully related---at least metaphorically or structurally---to the abstract framework of SLT, and in particular to the coherent state $|\alpha\rangle$ as defined in equation~(\ref{cstate}). While these connections are currently indirect and often heuristic in nature, our broader aim is to invite researchers in cognitive and sociopolitical sciences to engage with the QLM paradigm. We anticipate that future work will establish more formal, and ideally mathematical, correspondences between these domains.
                       
This interdisciplinary coupling opens new research avenues, including empirical estimation of social coherence parameters, simulation of social laser effects based on frame diffusion, and analysis of movement success in terms of resonance strength and social energy thresholds.

\subsection{Emotional Contagion and Affective Synchronization}

Emotions frequently spread across social groups, creating synchronized affective states that enhance collective readiness for coordinated action. A quantum-like model (QLM) emphasizing the dual-track nature of emotional processing - cognitive versus affective - was proposed in \cite{emotion}.

Hatfield et al. \cite{Hatfield}, in their seminal work \textit{Emotional Contagion}, argue that people often ``catch'' the emotions of others unconsciously. This occurs through automatic mimicry (copying facial expressions, vocal tones, or postures), feedback loops that reinforce emotions, and physiological coupling. Such mechanisms promote convergence in emotional states, fostering social cohesion, empathy, and coordinated behavior.  

\subsubsection*{Mapping to Social Laser Theory}

From the perspective of Social Laser Theory (SLT), automatic mimicry corresponds to phase alignment among social agents, feedback loops amplify emotional signals, and physiological coupling produces coherence in group affective states. Social cohesion lowers activation thresholds, while emotion propagation mirrors stimulated emission in the social lasing process.

Rim\'{e} \cite{R2009} extends this perspective, highlighting that emotions naturally drive social sharing. Individuals communicate their experiences almost immediately through verbal expression and repeated narration. Rim\'{e} distinguishes primary sharing (direct communication), secondary sharing (listeners retransmitting emotions), and tertiary sharing (broader social diffusion). Empirical studies suggest that 85--95\% of emotional experiences are shared, often engaging listeners emotionally. Cross-cultural research \cite{rime1991social} confirms this pattern across Belgium, Turkey, and Japan, with sharing typically peaking within 48 hours \cite{christophe1997social}. Experimental studies indicate that sharing enhances cognitive understanding but may not reduce emotional intensity, supporting QLM's dual-track framework \cite{zech2005does}. Secondary and tertiary sharing produce cascading flows of emotional information \cite{luminet2000social}, while large-scale events, such as the death of Princess Diana or the 9/11 attacks, demonstrate mass emotional alignment, described as ``collective effervescence.''

In SLT terms, primary and secondary sharing act as stimulated emotional emission, repeated narration pumps emotional energy, widespread sharing aligns phases across individuals, listener resonance strengthens coherence, and propagation cascades trigger large-scale activation.

\subsection{Framing Theory and Discursive Alignment}

Framing theory explains how social movements craft and communicate messages to mobilize supporters. A frame is an interpretive structure that helps people make sense of events and identities. Movements construct frames by defining problems, attributing causes, proposing solutions, and motivating action, while audiences bring pre-existing frames shaped by culture and experience. Mobilization is most effective when movement frames resonate with audience frames - a phenomenon known as \emph{frame resonance}.

Khrennikov's research on contextual decision-making intersects with framing processes \cite{AGS,framingP}, and Pothos and Busemeyer \cite{framingPB} explicitly discuss framing effects in cognitive contexts such as preference reversals.

Snow and Benford \cite{snow1988ideology} introduced frame resonance, emphasizing that effective framing combines a compelling narrative with deep cultural and emotional connection. This involves diagnostic framing (problem identification), prognostic framing (solution proposal), and motivational framing (rationale for action). Tarrow \cite{tarrow1998power} stresses that framing is dynamic, shaped by negotiation among movements, opponents, and political institutions.

Recent studies highlight the dynamic and emotional nature of frame alignment:

\begin{enumerate}[noitemsep]
    \item \textbf{Dynamic Alignment:} Frame resonance evolves with cultural and political context. Benford and Snow \cite{benford2000framing} identify bridging, amplification, extension, and transformation mechanisms.
    \item \textbf{Emotional and Moral Appeals:} Emotions such as anger, pride, and hope enhance resonance and motivate participation \cite{jasper2011emotions}.
    \item \textbf{Media and Digital Effects:} Social media accelerates frame diffusion and enables micro-targeted messaging.
    \item \textbf{Intersectional Framing:} Addressing multiple overlapping identities broadens appeal and mobilizes marginalized groups \cite{cho2013toward}.
\end{enumerate}

\subsubsection*{Coupling Framing Theory with Social Laser Theory}

Frame resonance corresponds to coherence in social energy states: just as lasers require phase coherence and population inversion, movements require aligned cultural frames for large-scale action. Discursive alignment synchronizes cognitive states, increasing the likelihood of ``social lasing'' - rapid, amplified mobilization. Threshold activation in SLT parallels motivational framing, where individuals shift from passive agreement to active participation. Iterative frame alignment mirrors SLT feedback loops, where social energy fluctuates due to internal discourse and external influences, including counter-framing and media dynamics.

Integrating these theories provides a rich toolkit for understanding social movements. Framing offers contextual socio-cultural and narrative grounding, while SLT models amplification and coherence dynamics. This combination enables research on social coherence parameters, simulation of social laser effects from frame diffusion, and analysis of movement success in terms of resonance strength and social energy thresholds (see also article \cite{framingP} for applications to finance).

\subsection{Collective Attention and Synchronization of Timelines}

In the digital age, collective attention often exhibits rapid, large-scale synchronization across populations, particularly during media events, crises, or breaking news. This synchronization manifests as temporally aligned bursts of online activity---especially on platforms like Twitter---where timelines converge around trending topics. These attention spikes reflect underlying cognitive processes, where millions of individuals orient their perception, emotion, and information processing around a common focus point, effectively creating a shared temporal and semantic frame.

Wu et al.\ \cite{wu2011who} provide foundational insights into how information diffuses on Twitter by analyzing the ``who says what to whom'' model of communication. Their study reveals that while most content remains within small networks, high-profile media events or celebrity communications can create bursts of attention that transcend typical network boundaries. This results in sudden synchronization of large numbers of users' timelines, indicating collective attention events.

Building on this, Lehmann et al.\ \cite{lehmann2012dynamical} identify distinct \textit{dynamical classes} of collective attention on Twitter. They classify attention patterns into categories such as ``sustained interest,'' ``transient bursts,'' and ``cyclical reactivations.'' Importantly, these patterns correspond to different kinds of events and topics---for instance, scheduled television events versus unexpected disasters---each producing a characteristic time-evolution of attention and synchronization among users. The emergence of these classes underscores how collective cognition is temporally structured and modulated by external stimuli, media practices, and social amplification mechanisms.

Together, these works suggest that the synchronization of timelines---observable in tweet volumes, retweet cascades, or hashtag use---serves as a real-time proxy for collective cognitive engagement. Such synchronization may also play a critical role in phenomena like narrative formation, belief propagation, and even protest mobilization, providing a fertile ground for modeling tools like Social Laser Theory or quantum-like attention dynamics.

\subsection{Resonance in Political Communication}

Political communication becomes particularly effective when messages \emph{resonate}---that is, align affectively and cognitively with the preexisting concerns, identities, or emotional states of recipients. This idea is central to social identity accounts of populist appeal, such as those developed by Mols and Jetten \cite{mols2016explaining}, who argue that right-wing populist messages gain traction not necessarily due to economic hardship, but because they affirm and activate perceived status threats among strongly identifying majority group members.

To formalize this concept of \emph{resonance} in QLM, we  use the formalism of complex Hilbert spaces ${\cal H}.$ Let $\psi_i$ represent the cognitive-emotional state of individual $i$, encoded as a normalized vector $\psi_i \rangle$ in ${\cal H}.$ Let $\hat{M}$ be a Hermitian positively semi-definite operator acting in ${\cal H}$ and representing the ideological and affective structure of a political message $m$, including its symbolic markers, emotional tone, and identity relevance. We define the resonance function $R_i(m)$ as the expectation value of $\hat{M}$ in state $\psi_i$:
\begin{equation}
\label{R}
R_i(m) = \langle \psi_i | \hat{M} | \psi_i \rangle .
\end{equation}
This formulation is not present in Mols and Jetten's original empirical study \cite{mols2016explaining}, but provides a QL abstraction to capture how certain messages selectively amplify alignment with specific cognitive-emotional configurations. The scalar product reflects the degree of alignment (or ``resonance'') between an individual's internal state  $| \psi_i \rangle$ and the projected content structure of a message - described by operator $M,$ enabling the modeling of selective susceptibility and communicative impact within populations.

In QLM, high $R_i(m)$ implies constructive interference between the message and individual's cognitive basis, resulting in greater amplitude for subsequent belief update or action:
\[
\text{Pr}(\text{activation} \mid m) \propto |R_i(m)|^2
\]

In this framework, resonance acts as a filter for message coherence across the population. When many individuals share similar cognitive basis states (e.g., via identity or ideology), high-fidelity resonance can emerge, analogous to phase coherence in wave systems. This can trigger large-scale opinion cascades, polarization, or rapid mobilization.

\subsubsection*{Integration with Social Laser Theory}

Formula (\ref{R}) can be used as a starting point towards integration of theoretical and practical studeis on 
resonance in political communication and SLT.

In SLT, resonant messages function as \emph{stimulated emissions}, aligning phase markers across individuals and facilitating coherent social action. If $\phi_i$ denotes the phase of an individual's ideological alignment and $\phi_m$ the phase encoded in the message, then the collective coherence can be modeled as:
\[
C(t) = \left| \frac{1}{N} \sum_{i=1}^N e^{i(\phi_i - \phi_m)} \right|
\]
High $C(t)$ corresponds to macroscopic coherence and readiness for synchronized action, especially if social energy levels have been sufficiently pumped.

\section{Integrative Table of Analogies}

\begin{center}
\begin{tabular}{|l|l|l|}
\hline
\textbf{Social Laser Concept} & \textbf{Analog} & \textbf{Observable} \\
\hline
Phase-alignment of agents & Frame resonance, emotional contagion & Message uptake, protest participation \\
Stimulated emission & Viral sharing, retweet cascades & Content virality metrics \\
Threshold crossing & Granovetter thresholds & Participation spikes \\
Coherent behavior & Flash mobs, mass rituals & Temporal/spatial coordination \\
Discourse pumping & Media agenda-setting & Repetition frequency, sentiment buildup \\
\hline
\end{tabular}
\end{center}

\section{From Conceptual Scaffolding Further on to Empirically Informed Theory Building} 

In the social sciences (that includes economics and management science) theories are key. They inform and guide the formation and formulation of research questions and research designs, and as a consequence thereof theoretical/mathematical modelling, as well as empirical and experimental modelling.  

The new normal in social sciences are not old system theories, but interaction theories, i.e. evolutionary contextual communication-based social (economic, cultural, etc.) interaction theories. We may also understand them as force-based theories, as opposed to boundary- or demarcation-based system theories.  

In old/classical social science, Newtonian dynamics are modelled along a completely deterministic approach. Now, with new paradigm of quantum social science on the horizon, we can see that the old entirely deterministic approach/theories are insufficient and obsolete. The 2007/2008 world financial and economic crisis in economics could not be predicted with classical models, as experts and economists using old (neoclassical) theory models were left dumbfounded, out in the cold. Neoclassical theories and models never were close to the pulse and reality of empirical markets, and of course never close to the empirical reality of human action in economics \cite{Mises}. The new dominant rising theory paradigm is quantum social science, and it also includes psychology and all behavioral sciences, i.e. it extends to all life sciences. 

With the new quantum paradigm, multitudes of causal forces rule, plus evolution and history of causalities are fully factored in, built on. Context is key, and not fully ignored (as e.g. in neoclassical economics). Social and life sciences are complex sciences, not simple sciences and not simple research topics that can be modelled with and after old Newtonian fully (100 percent) pre-deterministic laws/ways. The new normal and the new rational choice is quantum social science, the old weird choice is Newtonian-based classical social science. The new normal in all the social sciences (and life sciences) is evolutionary and fully contextual, it caters to the reality of non-commutative probabilities of psychological, social, economic, political and cultural events, series of events and developments.  

To make one united, i.e. universal theory, is not easy and takes time. All truly scientific theories have to be universal theories, i.e. universally applicable - otherwise they are not theories in the scientific sense. Scientific theories have to explain all events and developments within certain ranges/classes of events, cases and developments. There are `meta theories' that inform other `grand theories', and these again may inform and guide the set up and development of macro, meso and micro theories, or complement the very same. SLT, for example, is a grand theory that bridges both micro theory realms and macro theory ones.  

Quantum social science, and with it quantum-like modelling, embodies the much more rational approaches that are now (and have been since the last couple of decades) available to the broad masses of scientists and researchers. The genie is out of the bottle. It cannot be ignored, it cannot be unlearned, and it influences, informs, and leads to new progress in the decades ahead in every more research areas, topics and applications \cite{Aerts1,Aerts3,Aspalter1,Aspalter2}. 

The social and economic worlds, the cultural and political worlds are ruled by the principles of quantum social science, this is obvious, and this has been proven already by e.g. Aerts et al. \cite{Aerts1}and Orrell \cite{Orrell}, and many others. Aerts et al. \cite{Aerts1} made clear that - universally speaking - quantum-type behavior is valid in human language. Language is the foundation and driving force of every human cognition and decision-making and hence, of course, all human behavior and action. In quantum finance, Orrell \cite{Orrell} empirically tested and corroborated the supremacy of quantum-like modelling using large datasets, and with it demonstrated the non-empiricality of traditional classical modelling technigues that time and again failed to predict financial crises.  

The empirical evidence proving and/or supporting the supremacy of quantum-like modelling across the board is striking. The relative lack of and/or relatively slow scientific response can only be explained by Thomas Kuhn's `normal science' theory, where out-of-date theories rule, and where new innovative, successfully empirically tested theories are largely ignored for a long time \cite{Kuhn}. 

The prime research field in quantum social/life sciences is communication, intrapersonal and interpersonal communication, and everything that functions as and/or everything can be expressed with communication. This includes all human behaviors (incl. cognition and decision-making) and human actions in general \cite{Aspalter2,Aspalter4}. The smallest units of communications, i.e. words, are the quanta of society, economy and politics. Words, including numbers and symbols, in communication are the `communication quanta' of human cognition and decision-making, human behavior in particular and in general. As Aspalter \cite{Aspalter3} has put it, words in communication are the `least common denominators' that enable `universal quantum gates', without which universal theory-making in the social, economic, political etc. realm would be impossible. Without universal quantum gates, quantum computing mechanisms and quantum-like computing mechanisms in general would be impossible as well. These universal theoretical least common denominators combine all disciplines in the social and life sciences, they unite all theories and research fields. Aspalter \cite{Aspalter3} pointed out that all human-to-machine communications and vice versa, as well as all machine-to-machine communications are also part of the theoretical descriptive and explanatory plane of quantum social science, i.e. geometrically speaking. All of these communications and all the actors of human communications (human themselves, or their creations, like ideologies, ideas, markets and institutions) can be depicted as points in Hilbert spaces. To be clear, all social, economic and political actors/agents, including social movements and governments, can be depicted as points in Hilbert spaces \cite{Qadir}. 

Future directions in quantum social science research are manifold. The demand and opportunities are sheer endless. The development of different types of theories (positive theories including descriptive, explanatory and predictive theories, as well as normative theories) at different levels (meta, macro, meso, micro, or micro-cum-macro) needs all hands on deck. A new research community has been established in the past couple of decades, and is now taking off in terms of theory development, in terms of membership and in terms of modelling techniques and applications \cite{Aerts3,H&K,AAR,AGS,KHR1,laser5,KHR3,P&B,Orrell}.

\section{Further Research Directions} 
\label{FRD}

\begin{enumerate}[noitemsep] 

    \item \textbf{Empirical Study}: Use frequency and sentiment analysis of online discourse to track alignment and coherence over time. 

    \item \textbf{Theoretical Modeling}: Continue to construct a formal correspondence between concepts of coherence in quantum physics and frame resonance in political communication. 

    \item \textbf{Case Study}: Application of SLT to a real-world movement (e.g., Arab Spring, BLM), modeling activation thresholds, discursive pumping, and collective amplification. 

    \item \textbf{Simulation}: Development agent-based models with oscillatory internal states and simulate collective phase-locking and social cascades. 

\end{enumerate} 

SLT opens a pathway toward predictive modeling of social mobilization by providing a formal structure to: 

\begin{itemize} 

  \item Measure social energy distributions across populations. 

  \item Identify key mental markers driving resonance (first of all affective markers). 

  \item Estimate thresholds for coherence-induced phase transitions. 

  \item Simulate the dynamics of infon injection and collective response. 

  \item Develop further ``universally applicable theories'' of quantum behavior of people and in policy making, as well as quantum dynamics in and of social systems, markets and institutions. 

\end{itemize} 

 While still under development, SLT integrates symbolic, emotional, and informational dimensions into a unified model, promising new insights into the nonlinear dynamics of social change.

\section{Concluding Remarks: Towards a Neuro-Cognitive Theory of Collective Resonance}

Social Laser Theory (SLT) provides a powerful abstraction for modeling the emergence of large-scale social activation by applying concepts inspired by quantum optics---specifically coherence, excitation thresholds, and stimulated emission---to the human cognitive architecture. By formalizing these analogies, SLT offers a vital bridge between the individual neuro-cognitive state and collective dynamics. This framework highlights that mass mobilization is not merely an aggregation of classical preferences, but a result of phase alignment and the coherent synchronization of mental markers across a population.

The findings presented here suggest that cognitive state transitions, whether at the individual or institutional level, are inherently non-commutative, contextual, and stochastic. Policy evolution and institutional shifts follow a quantum-like trajectory where interference mechanisms---both constructive and destructive---dictate the outcome of decision-making processes. Within this ``Social Quantum Field,'' \textit{human entanglements} provide the invisible grid that creates, shapes, and upholds collective action. These entanglements facilitate culture and political hegemonies, being communicationally or environmentally entangled \cite{Aspalter2}. Consequently, everything in social science is contextual, mirroring the requirements of quantum mechanics wherever multiple social agents and informational quanta are involved.

By utilizing SLT, sociopolitical and economic case studies can be analyzed through a new lens of ``state changes'' and ``phase changes.'' The transition of social energy, driven by informational ``infons'' and institutional practices, can be translated into predictive models for policy shifts and economic actions \cite{Aspalter4}. This allows for a deeper understanding of sudden mass attitude-cum-behavior changes, ranging from public health compliance to shifts in investment patterns and financial market changes. 

Crucially, this article has established a foundational coupling between Quantum-Like Modeling (QLM) and key sociopolitical markers such as framing resonance and attention synchronization. While many correspondences remain structural, the theoretical architecture laid out here opens new avenues for transdisciplinary development across the social sciences and economics \cite{Conte, H&K, KHR3, AAR}. This shift moves the focus away from classical models of individual rationality \cite{Holcombe} toward a theory of collective coherence. It invites us to view sociopolitical change not as a simple linear progression, but as a nonlinear resonance phenomenon emerging from the repeated symbolic stimulation of mental markers and shared affective alignment.

Looking ahead, the development of SLT within the realm of neuroscience and cognitive science requires moving from formal scaffolds to empirical validation. We propose several tracks for future transdisciplinary research:
\begin{itemize}
    \item \textbf{Neuro-Cognitive Simulations:} Designing agent-based systems with oscillatory internal states to explore how phase coherence and infon injection trigger neuro-cognitive cascades;
    \item \textbf{Historical and Marker Dynamics:} Applying SLT to historical case studies to estimate resonance patterns, while formalizing the dynamics of cognitive and affective markers, emotional valence, and ideological phase as variables influencing infon absorption;
    \item \textbf{Predictive Coherence Tracking:} Developing indicators to monitor the buildup of coherence in online and physical discourses to anticipate ``tipping points'' or mass mobilization events;
    \item \textbf{Cross-Scale Validation:} Estimating activation thresholds and energy buildup in real-world social, economic, and political scenarios to forecast large-scale transformations.
\end{itemize}

In conclusion, SLT contributes a generalizable modeling scaffold that integrates the symbolic, affective, and neuro-informational components of behavior. By framing social change as a macroscopic manifestation of quantum-like effects across the human cognitive architecture, we open new directions for forecasting the nonlinear dynamics of our increasingly interconnected and reactive global society.

{\bf Acknowledgements:} For A. Kh. this research was partially supported by siting fellowships at Institute for Quantum Life Science (Chiba), RIKEN, Keio University, and Waseda University; A. Kh. would like to than profs. Makiko Yamada, Kazuhiro Sakurada, Yukio Guinji for scientific cooperation and hospitality.   

\section*{Appendix: Foundational Discussion}
 
Quantum theory offers a powerful and remarkably general mathematical framework for the rigorous analysis of quantum phenomena. However, as Niels Bohr emphasized early on, the theory does not describe the intrinsic behavior of quantum systems. Instead, it models experimental contexts and predicts, with extraordinary precision, the probability distributions of measurement outcomes.

According to the Copenhagen interpretation, as formulated by Bohr, quantum theory concerns itself with preparation and measurement procedures, and with the probabilistic relationship between them. Not all physicists accept this perspective, however. Quantum theory is characterized by a vast diversity of interpretations - even the Copenhagen interpretation itself has numerous ``sub-interpretations'' (and Plotnitsky has suggested referring instead to interpretations in the spirit of Copenhagen \cite{PL1}--\cite{PL3}). Engaging in debates over interpretation often opens a philosophical Pandora's box, and so we shall not pursue that path further here.

Our aim is simply to underline a widely held view: that quantum theory is not intended to ``explain'' phenomena in a classical sense, but rather to provide a calculational tool. The sentiment often attributed to Feynman - ``Shut up and calculate!'' - aptly captures this pragmatic stance, especially in response to demands for deeper explanations of core postulates such as Born's rule for computing quantum probabilities.

Some in the quantum community have nonetheless sought such explanations, particularly through the development of hidden variable theories. Using the Copenhagen interpretation in Bohr's spirit  \cite{PL1}--\cite{PL3} does not preclude the possibility of constructing such theories - and we will return to this issue later. But it is important to note that introducing hidden variables goes beyond conventional quantum theory. Typically, hidden variable theories are more complex than standard quantum mechanics, which is remarkably elegant and tractable, largely due to its linear mathematical structure.

Our purpose here is to advocate for the use of conventional quantum theory beyond its traditional domain of physics - especially in the social sciences - through the further development of SLT. Still, foundational issues remain central to our approach in QLM, because the systems under consideration are macroscopic and can, in principle, be described classically. In that sense, one might say that in QLM, hidden variables do exist.

Similarly to quantum modeling of physical processes, QLM provides a powerful and remarkably general apparatus for rigorous mathematical analysis of cognitive, sociopolitical phenomena (see above citations).  Yet, QLM falls short in offering detailed explanations of these processes in terms that resonate with cognition and sociopolitical sciences. In this regard, the situation mirrors that of quantum physics. 

In philosophic terms quantum theory is an epistemic theory - one that describes what can be known - formulated in terms of a peculiar and highly restricted class of observables. It is, of course, possible (though by no means necessary) to pursue ontic, or realist, theories that lie beneath quantum formalism - so-called hidden variables models. In the mainstream discourse of physics, it is often asserted - under assumptions considered ``natural'' - that such ontic theories are either nonexistent or physically implausible. Those that do exist are frequently dismissed as too exotic, even unphysical: Bohmian mechanics, with its nonlocality, or Nelson's stochastic mechanics, with its diffusion foundations, are prime examples. However, these sweeping no-go conclusions stem from an overly rigid and narrow framework governing the relationship between ontic and epistemic levels of theory - a framework primarily shaped by von Neumann, Bell, and Kochen-Specker.

An alternative, more generous framework is offered by {\it the Bild conception of theory} -- after Hertz, Boltzmann, Schr\"odinger (see \cite{KHRCambridge}  for review and reinterpretation) and more recently, Atmanspacher and Primas \cite{ATM}). The Bild framework loosens the straitjacket traditionally imposed on the prequantum$\to$quantum correspondence. Within this paradigm, one may construct hidden-variable models that underpin quantum theory without the naive identification of quantum observables with prequantum variables, as Bell once attempted. A key insight of the Bild approach is that the mapping from prequantum to quantum states is not one-to-one. Many distinct prequantum configurations may project onto the same quantum state - compare, for instance, with't Hooft's views on deterministic substrata \cite{Elze}. From this perspective, quantum theory offers not a mirror of reality but a blurred, approximate image - an echo rather than a voice.

That is why it is so important to also develop sociological and economic quantum theories from ground up. They are not limited by physical (or physically possible/technologically possible) boundaries. Evidence in social and economic sciences is there, it is plentiful everywhere. The problem is the opposite as in quantum physics. In social science, we have too much data, and too many variables, and this makes social science, including economics, so difficult (for those who perceive it as a science, as any other). In physics terminology, in the social sciences the degrees of freedom are often too numerous. But, not so in the case of Social Laser Theory, as it applies to special mass-scale social-psychological and political situations that (on purpose) lose details. People are losing their individuality, scores of individual traits, in social lasing events. That is, in Social Laser events only a few variables are guiding and triggering the social lasing process, effervescent mass reactions. 

We are indeed in need of many more social quantum theories. As with SLT, it is common in social sciences to work on a theory for decades. Some theories get defeated and others succeed in a competitive manner, where the all-deciding difference between defeat and success is always empirical evidence, and practicability in terms of applications and their outcomes. 

In QSFT building, including quantum economics theory building, theorists are not bound by traditional or widespread believes (as they are common also in physics). Quantum social science is more flexible, as we use quantum principles and quantum physical ontology is not the only base and are not the only building blocks to work with. We must also use all practical sociological, psychological and economic theories, and then integrate quantum principles or wisdom, quantum formalism and theory wherever it brings an advantage, wherever it leads to breakthroughs and innovation.

In social quantum theory building, human entanglements are replacing quantum-level entanglements. Weak time symmetry in the quantum realm is translated into echoes of the human past, and projections of expectations and evaluations of the future. Stories, emotions, logic and experiences of the past and stories/expectations of the future both become manifestations of social realities that are also alive and causally active in the here and now \cite{Aspalter4}. 

In SQFT, cultural predispositions, lingual connotations, emotional and thought-to-be-logical representations all become manifestations of the past in the here and now. The past is alive and the future is alive, too, both of which are constantly and vehemently being activated, changed and re-edited in the now. Socio-cultural and socio-political environmental changes, present-day education contents, media reports, social media contents are all triggering and re-editing these manifestations, which are causal manifestations in the sense of human entanglements. There are e.g. power exertions or Foucauldian power relations of the past activated and/or amplified (or dampened) in the here and now. The activation keys of today do not contain the action imperatives planted in the past, grown and constantly re-edited from the ancient past, millennia and centuries ago, to the current day. They are emergent. Also events themselves, or news about the events themselves, can and do trigger mass responses, shifts in mass sentiments, attitudes and mass action protocols, the likelihood of all choices - of thoughts, feelings and actions \cite{Aspalter3}. SLT has focused on such mass events and mass dynamics.

The limits and paradoxes of quantum physics itself translate into quantum social science problems (e.g. the need to explain and to refer to specific and/or all interpretations of quantum physics). But additional choices in between different interpretations of quantum physics, and choices like to avoid or accept working with e.g. `weak time symmetry' concepts all make up for difficulties that arise from a large, and growing, number of different in-between-themselves-incoherent schools (interpretations) of quantum mechanical and quantum field theories.

\end{document}